\documentstyle[aps]{revtex}

\def\stackunder#1#2{\mathrel{\mathop{#2}\limits_{#1}}}

\def\sumj{\mathop{\sum\nolimits^\prime}\limits_{n=-\infty}^{\infty}}
\begin{document}
\begin{center}
{\large\bf Completeness principle and quantum field theory on
nonglobally hyperbolic spacetimes} 

\vskip12pt
Sergey V. Sushkov\footnote{Email address: sushkov@kspu.ksu.ras.ru} 

\vskip6pt
{\small\em Department of Mathematics, Kazan State Pedagogical University,

Mezhlauk 1 Street, Kazan 420021, Russia
}

\vskip12pt
\parbox{14cm}{\small
We analyse in details the problems which one faces trying to
quantize a scalar field on the spacelike cylinder being the
simple example of a spacetime with closed timelike curves.
Our analysis brings to light the fact that the usual set of
positive and negative frequency solutions of the field equation
turns out to be incomplete. The consequence of this fact is that
the usual formulation of quantum field theory breaks down on
such a spacetime.

We postulate the completeness principle and build on its basis
the modified quantization procedure.  As an example, the
Hadamard function and $\langle\phi^2\rangle$ for 
the scalar field on the spacelike cylinder are calculated. It is shown
that the ``naive'' method of images gives the same results of
calculation. 

\vskip12pt
{\normalsize PACS number(s): 04.20.Gz, 04.62.+v}
} 

\end{center}
\vskip12pt
\section{Introduction}
During the last few years there has been considerable interest
in ``time machine'' physics, i.e., in classical and quantum
dynamics on nonglobally hyperbolic spacetimes which contain
closed timelike curves.  One of the problems, which the concept
of a time machine entails, is that the usual formulation of
quantum field theory breaks down on these spacetimes and instead
the question (both mathematical and conceptual) arises: what it
might mean to quantize a field on a nonglobally hyperbolic
spacetime?

So far there is no full understanding of this problem. First
attempts to clarify it were undertaken by Kay. He, in the
framework of algebraic approach, proposed the concept of
F-locality and postulated the F-locality condition \cite{Kay}
which has to play a role of ``minimal necessary conditions''
\cite{KRW} for any quantum field theory. 
The basis, the F-locality condition is founded on, is a
supposition that ``the laws in the small should coincide with
the usual laws for quantum field theory on globally hyperbolic
spacetimes'' \cite{Kay}. 
Spacetimes for which there exists a field algebra satisfying
F-locality was called by Kay as F-quantum compatible. The
importance of F-locality condition is in that it turns out to be
very restrictive. In particular, the general results was
recently proved by Kay, Radzikowski and Wald \cite{KRW} showing
that no spacetime containing a compactly generated Cauchy
horizon can be F-quantum compatible. As a consequence, this
means that the F-locality principle is {\em inconsistent} with
the existence of spacetimes containing closed timelike curves
(see further discussion of this problem in Refs. \cite{discus}).
One may interpret it as evidence that time machines are
unphysical (see Ref. \cite{KRW}) and must therefore be removed
from physics in agreement with Hawking's chronology protection
conjecture \cite{Hawking}. On the other hand, the F-locality
principle itself could still be revised \cite{Krasnikov}, or one
could consider some alternative approaches for constructing
quantum field theory on nonglobally hyperbolic spacetimes. For
example, recently Li and Gott \cite{LiGott} used the ``naive''
method of images to calculate a vacuum polarization in the
region of Misner space, containing closed timelike curves.

In this work we analyse in details the problems which one faces
trying to apply the standard canonical quantization method
for fields on spacetimes with closed timelike curves, 
and build the modified quantization procedure based on the
completeness principle.

The units $c=\hbar=1$ are used through the paper.

\section{Globally and nonglobally hyperbolic spacetimes} 
First of all, let us briefly discuss the notions of the {\em global
hyperbolic} and {\em nonglobal hyperbolic} spacetimes (one could
find more details in \cite{HE}). Let $(M,g)$ be a spacetime with
a Lorentzian metric $g$. The metric determines the causal
structure of spacetime by determining which curves are
spacelike, which are null and which are timelike at each point.
Locally, the metric will lead to a well defined light cone just
as in Minkowski space. However, the global restrictions enforced
by the assumption of an everywhere Lorentzian metric are rather
weak and still permit a wide variety of more or less exotic
global causal behaviour. The important distinction for our
purposes is between the global hyperbolic and nonglobal
hyperbolic spacetimes. We shall define a globally hyperbolic
spacetime to be a time-orientable spacetime which possesses a
Cauchy surface. Here, we define a Cauchy surface to be a smooth
spacelike surface no two points of which can be joined by a
causal curve (i.e. by a timelike or null curve), but such that
every inextendible causal curve in the spacetime intersects it
precisely once. Note that once there is one Cauchy surface in
the spacetime, there will be many. In particular, there will 
exist many global choices of time coordinate (so a globally
hyperbolic spacetime always arises as the product of the real
line with some 3-manifold) whose constant time surfaces are all
Cauchy. 

A spacetime which does not possess a Cauchy surface will be
called nonglobally hyperbolic.  The simplest example of a
nonglobally hyperbolic spacetime, which we shall examine in the
next section, is the two dimensional spacelike cylinder.

\section{Quantized scalar field on the spacelike cylinder} 
Consider the Minkowski plane with the metric 
        \begin{equation}\label{2Dmetr}
        ds^2=-dt^2+dx^2.
        \end{equation}
Choose a strip $\{t\in(-\infty,\infty),x\in[0,a]\}$ on this
plane and assume that points on the bounds $\gamma^-$: $x=0$ and
$\gamma^+$: $x=a$ are to be identified, so that $(t,0)\equiv(t,a)$.
After this procedure we obtain a spacetime with the topology
$S^1\times R^1$, called the timelike cylinder ${\cal T}_2$.
Note that ${\cal T}_2$ possesses a Cauchy surface; for example,
a hypersurface $t=\rm const$ is Cauchy. So the timelike cylinder
is a globally hyperbolic spacetime. 

Choose now a strip $\{t\in[0,a],x\in(-\infty,\infty)\}$ and
`glue' its bounds, so that $(0,x)\equiv(a,x)$. We obtain again a
spacetime with the topology of cylinder: $S^1\times R^1$, which 
shall be called the {\em spacelike cylinder} ${\cal S}_2$. It is
obvious that the spacelike cylinder contains closed timelike
curves; for example, the timelike curves $x=\rm const$ are
closed. So ${\cal S}_2$ is a nonglobally hyperbolic spacetime.

The spacetimes of cylinders ${\cal T}_2$ and ${\cal S}_2$ could
be represented as factor spaces ${\cal T}_2=M/{\cal R_{\cal T}}$
and ${\cal S}_2=M/{\cal R_{\cal S}}$, respectively, where $M$ is
an universal covering space (the Minkowski plane), ${\cal
R_{\cal T}}$ and ${\cal R_{\cal S}}$ are the equivalence
relations 
        \begin{eqnarray}\label{equivrel}
        \nonumber
        &&{\cal R_{\cal T}}:\quad (t,x+a)\equiv(t,x),\\
        &&{\cal R_{\cal S}}:\quad (t+a,x)\equiv(t,x).
        \end{eqnarray}

Note that our two dimensional examples may be converted into
four dimensional examples e.g. by taking the product with a flat
two dimensional Euclidian space, so that 
${\cal T}_4={\cal T}_2\times R^2$ and 
${\cal S}_4={\cal S}_2\times R^2$.

Consider a scalar field on ${\cal S}_2$.
Let $\phi$ be a real massless scalar field with the
Lagrangian 
        \begin{equation}\label{lagr}
        {\cal L}= \frac12 \partial_{\mu}\phi\partial^{\mu}\phi.
        \end{equation}
The corresponding field equation in the metric (\ref{2Dmetr}) reads
        \begin{equation}\label{2Dwe}
        [\partial_t^2-\partial_x^2]\phi(t,x)=0,
        \end{equation}
where $\partial_t=\partial/\partial t$,
$\partial_x=\partial/\partial x$. In addition, it follows 
from the identification rule (\ref{equivrel}) and from the
quadric form of the Lagrangian (\ref{lagr}) that the scalar
field has to obey the periodic condition
        \begin{equation}\label{2Dpc}
        \phi(t+a,x)=\phi(t,x).
        \end{equation}
(For brevity we will not consider the antiperiodic condition
$\phi(t+a,x)=-\phi(t,x)$, which is possible as well.)
It is easy to see that the functions
        \begin{equation}\label{2Dsols}
        e^{i k_n(x \pm t)}, 
        \end{equation}
where
        $$
        k_n=\frac{2\pi n}{a},\quad n=\pm1,\pm2,\dots,
        $$
form the set of particular solutions of the Eq.~(\ref{2Dwe}) obeying
the condition (\ref{2Dpc}).

Discuss now the problem of quantization of the scalar field on
${\cal S}_2$. In order to understand the essence of problem
better, we shall first of all try to apply straightforwardly the
standard scheme of the canonical quantization (we may refer, for
example, to \cite{BD} for more details of the canonical
quantization method). The main point of this approach is that a
spacetime is manifestly divided into space and time. In the
other words, the spacetime is foliated into spacelike
hypersurfaces labelled by a constant value of the time
coordinate $t$. Let $\Sigma$ be a particular spacelike
hypersurface $t={\rm const}$ with unit timelike normal vector
$n^\mu=(\frac{\partial}{\partial t})^\mu=(1,0)$.  The canonical
momentum $\pi$ conjugated with $\phi$ is defined by 
        $$
        \pi=\frac{\delta{\cal L}}{\delta(\partial_t\phi)}.
        $$
The canonical quantization means that the field $\phi$ is
regarded as an operator obeying the canonical commutation
relations 
        \begin{eqnarray}
        && [\phi(t,x),\phi(t,x')]=0, \nonumber \\
        && [\pi(t,x),\pi(t,x')]=0, \nonumber \\
        && [\phi(t,x),\pi(t,x')]=i\delta(x,x'),
        \end{eqnarray}
where the commutators are calculated in a given spacelike
hypersurface $t={\rm const}$ and $\delta(x,x')$ is a delta
function with the property that 
        \begin{equation}
        \int\limits_{-\infty}^{\infty}\delta(x,x')dx=1.
        \end{equation}

Now let us carry out the second quantization. With this aim we
should construct an orthonormal set of positive and negative
frequency modes being complete in a given spacelike hypersurface
$t=\rm const$, where a solution is defined to be positive
frequency with respect to the timelike normal vector $n^\mu$ if
$n^\mu\partial_\mu\phi=\partial_t\phi=-i\omega\phi$, $\omega>0$.
As usually, we divide the solutions
(\ref{2Dsols}) into positive and negative frequency modes
$\phi_n$ and $\phi_n^*$, respectively, where 
        \begin{equation}\label{phi_n}
        \phi_n=C_n e^{-i\omega t+i k_n x}, \quad \omega=|k_n|.
        \end{equation}
The modes $\phi_n$ and $\phi_n^*$ are orthonormal provided
$C_n=(4\pi\omega)^{-1/2}$, so that
        \begin{eqnarray}
        && (\phi_n,\phi_{n'})=\delta(k_n,k_{n'}), \nonumber \\
        && (\phi_n^*,\phi_{n'}^*)=-\delta(k_n,k_{n'}), \nonumber \\
        && (\phi_n^*,\phi_{n'})=0,
        \end{eqnarray}
where the scalar product of a pair of solutions $f_1$ and $f_2$
is defined as
        \begin{equation}
        (f_1,f_2)=-i\int\limits_{-\infty}^{\infty}dx\big(f_1\partial_t f_2^*-
        \partial_tf_1\, f_2^*\big)_{t={\rm const}}.
        \end{equation}
Unfortunately, the set of modes $\{\phi_n,\phi_n^*\}$ is not
possessing the important property; namely, it is {\em
incomplete} in a given spacelike hypersurface $t=\rm const$. This
means that an arbitrary solution $\phi(t,x)$
cannot be represented as an expansion in terms of positive and
negative frequency modes. To make sure of the incompleteness of
the set $\{\phi_n,\phi_n^*\}$ it is enough to note that the
necessary condition of completeness is not satisfied, i.e., 
        \begin{equation}
        \frac{1}{a}\sum_{n=-\infty}^{\infty} e^{ik_n(x-x')}=
        \sum_{n=-\infty}^{\infty}\delta(x-x'-na) \not= \delta(x,x').
        \end{equation}
The incompleteness of the set of classical modes leads to the
fact that the field operator $\phi$ cannot be in general
represented as a sum $\sum_n(a_n\phi_n+a_n^\dagger\phi_n^*)$,
where $a_n$ and $a_n^\dagger$ are usual annihilation and
creation operators. As a result, a natural
definition of a vacuum state $|0\rangle$, such that
$a_n|0\rangle=0$, simply does {\em not} exist in this approach.
Thus, we must conclude that the standard scheme of the canonical
quantization turns out to be {\em inapplicable} and has to be
modified.  

The preceding analysis prompts to us at least two directions in which
we might modify the standard quantization procedure. Discuss briefly
first possibility. 

\subsection*{Local vacuum} 
Thus, we have found that the problems of quantum field theory on
nonglobally hyperbolic spacetimes are connected with the fact
that the set $\{\phi_n,\phi_n^*\}$ of usual positive and
negative frequency modes turns out to be incomplete in a given
spacelike hypersurface $\Sigma$, and so there does not exist an
expansion in terms of $\phi_n$ and $\phi_n^*$ for an arbitrary
solution $\phi(t,x)$.  Note that we mean a global expansion,
i.e., that whose coefficients do not depend on a point of the
hypersurface. At the same time, it is still possible to make
such an expansion {\em locally}, i.e., at each separate point
of $\Sigma$, so that
$f(t,x)|_\Sigma=\sum_n[A_n(x)\phi_n+B_n(x)\phi_n^*]$, where the
coefficients $A_n(x)$ and $B_n(x)$ are functions of a point of
$\Sigma$. Let us stress that such the expansion has no
universal character and can be realized in many ways with
various coefficients $A_n(x)$ and $B_n(x)$. Nevertheless, assume
that some choice of $A_n(x)$ and $B_n(x)$ has been done (for
example, because of physical arguments). The corresponding field
operator $\phi$ could be now represented as a sum
$\sum_n[a_n(x)\phi_n+a_n^\dagger(x)\phi_n^* ]$, where $a_n(x)$
and $a_n^\dagger(x)$ are the local annihilation and creation
operators, respectively. A vacuum state $|0\rangle$, defined by
$a_n(x)|0\rangle=0$, $\forall n$, turns out to be depending on a
point of the spacelike hypersurface $\Sigma$. In the other
words, vacuum is in this case defined {\em locally}. 
This seems to be unsatisfactory with the physical
point of view.  Below we suggest another way to modify the
quantization procedure which is free from this shortcoming. 

\subsection*{The completeness principle} 
The completeness of a set of positive and negative frequency
modes is an {\em essential} property of quantum field theory
on a globally hyperbolic spacetime. We shall regard this
property to be a fundamental feature of any quantum field theory
and put it into the basis of the modified quantization procedure
postulating \\[6pt]
{\bf The completeness principle:}~{\em A spacetime has to be foliated 
into hypersurfaces (not necessarily spacelike) so that the
corresponding set of positive and negative frequency modes would
be complete.}

\vskip6pt
Let us apply this principle for the scalar field on the
spacelike cylinder ${\cal S}_2$. To construct a complete
orthonormal set of modes we choose a foliation of ${\cal S}_2$
into {\em timelike} hypersurfaces. Let $\Sigma$ be a particular
hypersurface $x={\rm const}$ with unit spacelike normal vector
$n^\mu=(\frac{\partial}{\partial x})^\mu=(0,1)$.  A canonical
momentum $\pi$ conjugated with $\phi$ should now be defined by 
        \begin{equation}
        \pi
        =\frac{\delta{\cal L}}{\delta(\partial_x\phi)}.
        \end{equation}

As usual, we shall say that a solution is positive frequency with
respect to the normal vector $n^\mu$ if
$n^\mu\partial_\mu\phi=\partial_x\phi=-i\kappa\phi$, $\kappa>0$. 
In our case the functions $f_n$ and $f_n^*$, where
        \begin{equation}\label{set}
        f_n=C_n e^{-i\kappa x+ik_n t}, \quad \kappa=|k_n|,
        \end{equation}
form the set $\{f_n,f_n^*\}$ of positive and negative frequency
modes, respectively. One may check that the modes are orthonormal 
provided $C_n=(2a\kappa)^{-1/2}$, so that
        \begin{equation}
        (f_n,f_{n'})=\delta_{nn'}, \quad
        (f_n^*,f_{n'}^*)=-\delta_{nn'}, \quad
        (f_n,f_{n'}^*)=0,
        \end{equation}
where the scalar product is defined as
        \begin{equation}
        (f_1,f_2)=-i\int\limits_{0}^{a}dt\big(f_1\partial_x f_2^*-
        \partial_xf_1\, f_2^*\big)_{x={\rm const}}.
        \end{equation}
(Note that the mode $n=0$ should be excluded from the set
$\{f_n,f_n^*\}$ because of non-normability.)
The set $\{f_n,f_n^*\}$ is {\em complete} on a
given timelike hypersurface $x=\rm const$. To make sure of this
we note that the necessary condition of completeness is
satisfied, i.e., 
        \begin{equation}
        \frac{1}{a}\sum_{n=-\infty}^{\infty} e^{ik_n(t-t')}=
        \sum_{n=-\infty}^{\infty}\delta(t-t'-na) = \delta(t,t'),
        \end{equation}
where $\delta(t,t')$ is a delta function with the property that
        \begin{equation}
        \int\limits_{0}^{a}\delta(t,t')dt=1.
        \end{equation}
To quantize the scalar field we shall assume that $\phi$
is an operator obeying the ``canonical'' commutation relations 
        \begin{eqnarray}\label{comrel}
        && [\phi(t,x),\phi(t',x)]=0, \nonumber \\
        && [\pi(t,x),\pi(t',x)]=0, \nonumber \\
        && [\phi(t,x),\pi(t',x)]=i\delta(t,t'),
        \end{eqnarray}
where the commutators are calculated on a given timelike
hypersurface $x=\rm const$.
The field operator $\phi(t,x)$ may be represented as an
expansion in terms of modes of the complete set $\{f_n,f_n^*\}$:  
        \begin{equation}\label{expansion}
        \phi(t,x)=\sum_n(b_n f_n+b_n^\dagger f_n^*)
        \end{equation}
Substituting this expansion into the Eqs.~(\ref{comrel}) gives
the commutation relations for operators $b_n$ and $b_n^\dagger$:
        \begin{equation}\label{bbcomrel}
        [b_n,b_{n'}]=0,\quad
        [b_n^\dagger,b_{n'}^\dagger]=0,\quad
        [b_n,b_{n'}^\dagger]=\delta_{nn'}.
        \end{equation}
The operators $b_n$ and $b_n^\dagger$ act on orthonormal
ket-vectors (Fock states) $|\ \rangle$ of Hilbert space $\cal
H$. A vacuum state $|0\rangle$ is defined by $a_n|0\rangle=0,\,
\forall n$, and describes the situation when no particles are in
a given timelike hypersurface $x=\rm const$. The operator
$b_n^\dagger$ determines a one-particle state $|1_n\rangle$ by
$|1_n\rangle=b_n^\dagger|0\rangle$. Similarly, one may construct
many-particle states (see e.g. Ref. \cite{BD}).

To make clear the physical sense of Fock states $|\ \rangle$
we consider the operator  
        \begin{equation}\label{Hamop}
        H=\int_\Sigma T_{\mu\nu}n^\mu n^\nu d\Sigma,
        \end{equation}
where $T_{\mu\nu}$ is the stress-energy tensor of the scalar
field, which could be obtained by the standard way in the
following form: 
        \begin{equation}
        T_{\mu\nu}=\partial_\mu\phi\partial_\nu\phi-
        \frac12 g_{\mu\nu}g^{\alpha\beta}
        \partial_\alpha\phi\partial_\beta\phi .
        \end{equation}
Note that the expression (\ref{Hamop}) determines 
Hamilton or {\em total energy} operator for a globally
hyperbolic spacetime with $\Sigma$ being a Cauchy surface. Which
sense get this operator in the considered case when ${\cal S}_2$
is a nonglobally hyperbolic spacetime and $\Sigma$ is a timelike
hypersurface? For a given hypersurface $x=\rm const$ we obtain
        \begin{equation}
        H=\int\limits_0^a T_{11} dt=
        \frac12\int\limits_0^a [(\partial_t\phi)^2+
        (\partial_x\phi)^2] dt.
        \end{equation}
Substituting $\phi$ from (\ref{expansion}) and taking into
account the relations (\ref{bbcomrel}) we find
        \begin{equation}\label{toten}
        H=\sum_n (b_n^\dagger b_n+\frac12) \kappa,
        \end{equation}
The operator $N_n\equiv b_n^\dagger b_n$ is a particle number
operator for mode $n$. Supposing that the energy of each quantum
in mode $n$ is equal to $\kappa$ we find that $H$, given by Eq.
(\ref{toten}), is a positive value total energy operator.

\section{Hadamard function and $\langle\phi^2\rangle$}
Consider now an example of calculation of a vacuum expectation
value in a nonglobally hyperbolic spacetime. In this section we
shall calculate the Hadamard function and the renormalized
vacuum expectation value of field square $\langle\phi^2\rangle$
for the scalar field on the four dimensional timelike cylinder
${\cal S}_4$.

The spacetime ${\cal S}_4$ may be represented as the factor
space $M/\cal R$, where $M$ is Minkowski spacetime with the
metric 
        \begin{equation}\label{metrfour}
        ds^2=-dt^2+dx^2+dy^2+dz^2,
        \end{equation}
and $\cal R$ is the equivalence relation
        \begin{equation}
        (t+a,x,y,z)\equiv (t,x,y,z).
        \end{equation}
The scalar field $\phi$ with the Lagrangian (\ref{lagr}) obeys
the field equation 
        \begin{equation}\label{fefour}
        [\partial_t^2-\partial_x^2-
        \partial_y^2-\partial_z^2]\phi=0
        \end{equation}
and, additionally, has to satisfy the periodic condition
        \begin{equation}\label{pcfour}
        \phi(t+a,x,y,z)=\phi(t,x,y,z).
        \end{equation}
The particular solutions of the field equation (\ref{fefour})
obeying the periodic condition (\ref{pcfour}) are
        \begin{equation}
        e^{\pm i\kappa x}e^{i(\omega_n t+k_y y+k_z z)},
        \end{equation}
where 
        \begin{equation}
        \omega_n=\frac{2\pi n}{a},\quad 
        \kappa=\sqrt{\omega_n^2-k_y^2-k_z^2},\quad
        n=\pm 1,\pm 2,\dots
        \end{equation}
Demanding $\kappa$ to be real gives
        \begin{equation}
        k_y^2+k_z^2 \leq \omega_n^2.
        \end{equation}

Choose a foliation of ${\cal S}_4$ into timelike hypersurfaces.
Let $\Sigma$ be a particular hypersurface 
$x={\rm const}$ with unit timelike normal vector
$n^\mu=(\frac{\partial}{\partial x})^\mu=(0,1,0,0)$, which
defines a positive frequency solution by
$n^\mu\partial_\mu\phi=\partial_x\phi=-i\kappa\phi$, $\kappa>0$.
The positive frequency solutions may be taken in the following
form: 
        \begin{equation}\label{4Dmodes}
        f_J=\frac{1}{\sqrt{8\pi^2 a\kappa}}e^{-i\kappa x}
        e^{i(\omega_n t+k_y y+k_z z)},
        \end{equation}
where $J\equiv\{n,k_y,k_z\}$. The functions $f_J$ and $f_J^*$ form
the set $\{f_J,f_J^*\}$ of positive and negative frequency 
modes, which are orthonormal with respect to the scalar product
        \begin{equation}
        (f_1,f_2)=-i\int\limits_0^a dt \int\!\!\!\int dy\,dz
        (f_1\partial_x f_2^* - \partial_x f_1\, f_2^*)_{x={\rm const}}.
        \end{equation}
The important property of the set $\{f_J,f_J^*\}$ is that it is
complete on a given timelike hypersurface $x=\rm const$.

The Hadamard function for the scalar field is defined as
        \begin{equation}\label{defHad}  
        G^{(1)}(X,X')=\langle0|\phi(X)\phi(X')+
        \phi(X')\phi(X)|0\rangle,
        \end{equation}
where $X=(t,{\bf r})=(t,x,y,z)$. Represent the field operator
$\phi(X)$ as an expansion in terms of modes of the set
$\{f_J,f_J^*\}$: 
        \begin{equation}
        \phi(X)=\sum_J(b_J f_J+b_J^\dagger f_J),
        \end{equation}
where 
        \begin{equation}
        \sum_J=\sumj
        \stackunder{k_y^2+k_z^2\leq\omega_n^2}
        {\int\!\!\!\int}dk_y\,dk_z
        \end{equation}
and the prime means that the mode $n=0$ is excluded. (Note that
the mode $n=0$ should be excluded because of non-normability.)
Substituting this expansion into the expression (\ref{defHad})
and taking into account that $b_J|0\rangle=0$, $\forall J$, we
obtain 
        \begin{equation}
        G^{(1)}(X,X')=\sum_J\left[f_J(X)f_J^*(X')+
        f_J^*(X)f_J(X')\right].
        \end{equation}
By using the modes (\ref{4Dmodes}) the Hadamard function may be
represented in the following form:
        \begin{equation}\label{Had}
        G^{(1)}_{\cal S}(X,X')=
        \frac{1}{4\pi^2 a}\sumj
        \stackunder{{k_y^2+k_z^2\leq\omega_n^2}}
        {\int\!\!\!\int}dk_y\,dk_z
        \frac1\kappa \cos(\kappa\triangle x-
        \omega_n\triangle t-k_y\triangle y-k_z\triangle z),
        \end{equation}
where $\triangle X=X-X'$ and the subscript ${\cal S}$ denotes
the Hadamard function on ${\cal S}_4$. 
Carrying out the appropriate transformations of the above
expression (see the appendix for more details) we finally obtain
        \begin{equation}\label{HadS}
        G_{\cal S}^{(1)}(X,X')=-\frac{1}{4\pi a\triangle r} 
        \left[
        \frac{\displaystyle\sin\frac{2\pi}{a}(\triangle t-\triangle r)} 
        {\displaystyle 1-\cos\frac{2\pi}{a}(\triangle t-\triangle r)}-
        \frac{\displaystyle\sin\frac{2\pi}{a}(\triangle t+\triangle r)} 
        {\displaystyle 1-\cos\frac{2\pi}{a}(\triangle t+\triangle r)}
        \right].
        \end{equation}
As usual the renormalized Hadamard function is taken to be 
        \begin{equation}
        G_{\cal S}^{(1)\rm ren}(X,X')=G_{\cal S}^{(1)}(X,X')-
        G_M^{(1)}(X,X'),
        \end{equation}
where $G_M^{(1)}(X,X')$ is the Hadamard function for the usual
Minkowski vacuum \cite{BD}:
        \begin{equation}\label{HadM}
        G_M^{(1)}(X,X')=
        -\frac{1}{2\pi^2}\frac1{(t-t')^2
        -(x-x')^2-(y-y')^2-(z-z')^2}.
        \end{equation}
The renormalized vacuum expectation value of the field square
$\langle\phi^2\rangle_{\cal S}$, characterizing vacuum
fluctuations on ${\cal S}_4$, is found as follows:
        \begin{equation}\label{phisquare}
        \langle\phi^2\rangle_{\cal S}=
        \lim_{X'\to X}G_{\cal S}^{(1)\rm ren}(X,X').
        \end{equation}
Carring out simple calculations we obtain
        \begin{equation}\label{phiS}
        \langle\phi^2\rangle_{\cal S}=-\frac1{6a^2}.
        \end{equation}

For comparison we write down the corresponding expression for
$\langle\phi^2\rangle$ calculated for the scalar
field on the `usual' spacelike cylinder ${\cal T}_4$ \cite{BD}:
        \begin{equation}\label{phiT}
        \langle\phi^2\rangle_{\cal T}=\frac1{6a^2}.
        \end{equation}
Note that the absolute values of $\langle\phi^2\rangle_{\cal S}$
and $\langle\phi^2\rangle_{\cal T}$ are equal, whereas their signs
are opposite.

\section{Method of images}
An example of a nonglobally hyperbolic spacetime is the Misner
space (see e.g. \cite{HE}). The two-dimensional Misner space may
be obtained by quotienting the region $t+x>0$ of Minkowski plane
by a fixed Lorentz boost. As a manifold, it is again a cylinder,
but the metric -- while locally flat -- differs globally from
either the timelike or the spacelike cylinder. In fact, the
cylinder is divided into two halves by a single closed null
geodesic (the surface labelled $\tau=0$) and one can show that
the open half cylinder for $\tau<0$ is conformally isometric to
the timelike cylinder, while the open half cylinder for $\tau>0$
is conformally isometric to the spacelike cylinder. In the other
words, the Misner space consists of a globally hyperbolic region
($\tau<0$) and a region of closed timelike curves ($\tau>0$)
separated by the surface $\tau=0$, which is called the
chronology horizon. A quantum field theory is well-defined in
the first region, and we may refer to the pioneering work by
Hiscock and Konkowski \cite{HK} and also to \cite{Sushkov},
where the vacuum polarization of a scalar field in the globally
hyperbolic region of the Misner space has been calculated and
discussed the behaviour of the renormalized stress-energy tensor
near the chronology horizon. At the same time, since there is
no well-defined quantum field theory in nonglobally hyperbolic
spacetimes, so until recently there was no calculations of
the vacuum polarization in the region of closed timelike curves
of the Misner space. However, recently Li and Gott \cite{LiGott}
represented such a calculation. They used {\em a priori} the
method of images adapted to the case of spacetimes with closed
timelike curves. 

Apply the method of images for calculation of the vacuum
polarization of the scalar field on ${\cal S}_4$.
As was mentioned, the universal covering space of ${\cal S}_4$
is Minkowski space. The Hadamard function $G_M^{(1)}(X,X')$ for
the usual Minkowski vacuum is given by Eq. (\ref{HadM}). Taking into
account the periodicity in time (\ref{pcfour}) we may construct
the Hadamard function on ${\cal S}_4$ as an image sum
        \begin{eqnarray}\label{HadImage}
        \nonumber&& \lefteqn{
        G_{\cal S}^{(1)}(t,{\bf r};t',{\bf r}')=
        \sum_{n=-\infty}^{\infty} 
        G_M^{(1)}(t,{\bf r};t'+na,{\bf r}')} \\
        && =-\frac1{2\pi^2}\sum_{n=-\infty}^{\infty} 
        \frac{1}{(t-t'+na)^2
        -(x-x')^2-(y-y')^2-(z-z')^2}.
        \end{eqnarray}
Using the formula (\ref{phisquare}), which gives the value of
$\langle\phi^2\rangle$, we obtain after simple calculations 
        \begin{equation}\label{phi2image}
        \langle\phi^2\rangle_{\cal S}=-\frac1{\pi^2 a^2}
        \sum_{n=1}^{\infty}\frac1{n^2}=-\frac1{6a^2}.
        \end{equation}
Note that we have obtained the result which coincides with Eq.
(\ref{phiS}). Thus, we may conclude that the method of images
is consistent with the modified quantization procedure based on
the completeness principle.

\section{Concluding remarks}
Summarizing, let us stress once more that the completeness
principle plays an essential role in quantum field theory on
nonglobally hyperbolic spacetimes. Mathematically, this simply
supposes the existence of a complete orthonormal basis of
solutions of classical field equations. In the paper we have
revealed the fact that in order to construct such a basis in a
spacetime with closed timelike curves one should consider a
foliation of spacetime into timelike hypersurfaces.  A given
initial timelike hypersurface performs a role like that of a
spacelike Cauchy surface in a globally hyperbolic spacetime. In
particular, the scalar product of a pair of solutions is
determined on the hypersurface, and the conjugate momentum of a
field is defined by a spacelike unit vector normal to the
hypersurface.

Note that, with the mathematical point of view, there is no
principal difference between a field theory on globally or
nonglobally hyperbolic spacetimes. One just supposes the initial
hypersurface to be spacelike (a Cauchy surface) or timelike,
respectively. However, the situation is quite different with the
physical point of view. Really, as is known in the case of
globally hyperbolic spacetimes one specifies field's initial
values on the Cauchy surface at a given moment of time, and this
does uniquely determine the following field's evolution in time.
But in the case of a spacetime with closed timelike curves one
should specify field's initial values on a timelike
hypersurface. As a result, the field becomes to be determined at
each moment of time, and hence there is {\em no} evolution
in this case. 
Of course, this situation is a direct consequence of the
existence of closed timelike curves. 

\section*{Acknowledgement}
This work was supported in part by the Russian Foundation of Basic
Research grant No 99-02-17941.

\section*{Appendix}
\setcounter{equation}{0}\renewcommand{\theequation}{A\arabic{equation}}
Consider the expression (\ref{Had}) for $G_{\cal S}^{(1)}$ and
carry out the following transformation:
        \begin{eqnarray}\label{A1}
        &&\frac{1}{4\pi^2 a}\sumj
        \stackunder{{k_y^2+k_z^2\leq\omega_n^2}}
        {\int\!\!\!\int}dk_y\,dk_z
        \frac1\kappa \cos(\kappa\triangle x-
        \omega_n\triangle t-k_y\triangle y-k_z\triangle z)
        \nonumber\\
        &&=\frac{1}{2\pi^2 a}\sum_{n=1}^{\infty}
        \cos\omega_n\triangle t
        \stackunder{{k_y^2+k_z^2\leq\omega_n^2}}
        {\int\!\!\!\int}dk_y\,dk_z
        \frac1\kappa 
        \cos(\kappa\triangle x-k_y\triangle y-k_z\triangle z).
        \end{eqnarray}
To calculate the integral in the last expression we make a
change of variables. Introduce the polar coordinates
$(k,\varphi)$ in the space of vectors ${\bf k}=(k_y,k_z)$, so
that 
        \begin{eqnarray}
        &&k_y=k\cos\varphi,\ k_z=k\sin\varphi, \nonumber\\
        &&k=\sqrt{k_y^2+k_z^2},\ dk_ydk_z=k\,dk\,d\varphi. \nonumber
        \end{eqnarray}
Assume also that the polar axis is directed along the vector
${\bf v}=(\triangle y,\triangle z)$, so that ${\bf
kv}=kv\cos\varphi$, where $v=\sqrt{\triangle y^2+\triangle z^2}$.
Now we may transform the expression (\ref{A1}) as follows:
        \begin{eqnarray}
        &&\frac{1}{2\pi^2 a}\sum_{n=1}^{\infty}
        \cos\omega_n\triangle t
        \int\limits_0^{\omega_n} \int\limits_0^{2\pi}
        \frac{k\,dk\,d\varphi}{\sqrt{\omega^2-k^2}}
        \cos(\sqrt{\omega^2-k^2}\triangle x-kv\cos\varphi) \nonumber\\ 
        &&=\frac{1}{\pi a}\sum_{n=1}^{\infty}
        \cos\omega_n\triangle t
        \int\limits_0^{\omega_n} 
        \frac{k\,dk}{\sqrt{\omega^2-k^2}}
        \cos\sqrt{\omega^2-k^2}\triangle x \, J_0(kv) \nonumber\\ 
        &&=\frac{1}{\pi a \triangle r}\sum_{n=1}^{\infty}
        \cos\omega_n\triangle t \, \sin\omega_n\triangle r \nonumber\\ 
        &&=-\frac{1}{2\pi a \triangle r}\sum_{n=1}^{\infty}\left[
        \sin\omega_n(\triangle t-\triangle r)-
        \sin\omega_n(\triangle t+\triangle r) \right],
        \end{eqnarray}
where $J_0(z)$ is the Bessel function and $\triangle
r=\sqrt{\triangle x^2+\triangle y^2+\triangle z^2}$.
To carry out the last transformations we used the following
formulas (see e.g. \cite[(2.5.40.10)]{PBMa} and
\cite[(2.12.21.6)]{PBMb}):
        $$
        \int\limits_{x_0}^{x_0+2\pi}e^{-iz\sin x}dx=2\pi J_0(z),
        $$
        $$
        \int\limits_0^a x\frac{\cos b\sqrt{a^2-x^2}}{\sqrt{a^2-x^2}} 
        J_0(cx)dx=\frac{\sin a\sqrt{b^2+c^2}}{\sqrt{b^2+c^2}}.
        $$
So, we obtain the next form of $G_{\cal S}^{(1)}$:
        \begin{equation}\label{A3}
        G^{(1)}_{\cal S}(X,X')=-\frac{1}{2\pi a\triangle r}
        \sum_{n=1}^{\infty}
        \big[\sin\omega_n(\triangle t-\triangle r)-
         \sin\omega_n(\triangle t+\triangle r)\big].
        \end{equation}
To calculate the series in (\ref{A3}) we use the
`trick' with regularization 
        $$
        \sum_{n=1}^{\infty}\sin nx=
        \lim_{\lambda\to 0}\sum_{n=1}^{\infty}e^{-n\lambda}\sin nx
        $$
and apply the following formula \cite{PBMa}:
        $$
        \sum_{n=1}^{\infty}e^{-n\lambda}\sin nx=
        \frac12\frac{\sin x}{\cosh\lambda-\cos x}.
        $$
Finally, we find
        \begin{equation}
        G_{\cal S}^{(1)}(X,X')=-\frac{1}{4\pi a\triangle r} 
        \left[
        \frac{\displaystyle\sin\frac{2\pi}{a}(\triangle t-\triangle r)} 
        {\displaystyle 1-\cos\frac{2\pi}{a}(\triangle t-\triangle r)}-
        \frac{\displaystyle\sin\frac{2\pi}{a}(\triangle t+\triangle r)} 
        {\displaystyle 1-\cos\frac{2\pi}{a}(\triangle t+\triangle r)}
        \right].
        \end{equation}

\end{document}